\documentclass{article}
\usepackage{spconf,amsmath,graphicx,hyperref}

\usepackage{amsfonts,amssymb}
\usepackage{booktabs}
\usepackage{etoolbox,siunitx}
\usepackage{cite}
\usepackage{mathtools}
\usepackage{cleveref}
\usepackage{IEEEtrantools}

\usepackage{bm}

\DeclareMathOperator*{\argmax}{arg\,max}
\usepackage{multirow,multicol}

\usepackage[table]{xcolor}
\usepackage{pgf}
\usepackage{collcell}

\newcommand*{\MinNumber}{18}
\newcommand*{\MaxNumber}{60}

\newcommand{\ApplyGradient}[1]{%
  \pgfmathsetmacro{\PercentColor}{100.0*(#1-\MinNumber)/(\MaxNumber-\MinNumber)}%
  \edef\x{\noexpand\cellcolor{red!\PercentColor}}\x\textcolor{black}{#1}%
}
\newcolumntype{R}{>{\collectcell\ApplyGradient}{r}<{\endcollectcell}}

\usepackage[absolute,overlay]{textpos}
\newcommand{\copyrightstatement}{
    \begin{textblock}{1}(0.108, 0.935)
         \noindent
         \footnotesize
         \copyright 2026 IEEE. Personal use of this material is permitted. Permission from IEEE must be obtained for all other uses, in any current or future media, including reprinting/republishing this material for advertising or promotional purposes, creating new collective works, for resale or redistribution to servers or lists, or reuse of any copyrighted component of this work in other works. DOI: \href{https://doi.org/10.1109/ICASSP55912.2026.11461520}{10.1109/ICASSP55912.2026.11461520
}
    \end{textblock}
}

\title{Single-Microphone Audio Point Source Discriminative Localization  From Reverberation Late Tail Estimation}
\name{Matthew Maciejewski}
\address{Human Language Technology Center of Excellence, Johns Hopkins University, Baltimore, USA}

\begin{document}
\copyrightstatement
\ninept

\bstctlcite{IEEEexample:BSTcontrol}

\maketitle

\begin{abstract}
Location information can be a valuable signal for audio segmentation tasks, especially as a complement to methods focusing on the content or qualities of the sources.
Though audio source localization is typically performed using the observations of the signal captured by multiple microphones in space, information about a source's location is captured by a single microphone through its arrival time and spectral amplitude---given the source's emitted signal is known.
Since reverberation originates from the audio sources in a room, it accordingly contains some information about the emitted audio signals.
The late-tail part of reverberation is relatively invariant to the local source and microphone geometry, depending primarily on only the room itself, and thus can provide the necessary reference information about audio signals that depends minimally on their location.
In this work, we leverage the robust late-tail estimation of Weighted Prediction Error~(WPE) dereverberation within a probabilistic framework to estimate the likelihood of two audio signals collected in the same room as having originated from the same location.
We demonstrate the effectiveness of our approach on the speaker diarization task in both simulated and real environments.
\end{abstract}


\section{Introduction}
\label{sec:introduction}
Outside of heavily-controlled settings like recording studios, audio recordings typically capture multiple audio signals emanating from multiple sources.
Many audio technologies then require segmentation of audio recordings: marking the temporal regions containing the audio coming from different sources of interest, e.g. sound event detection~\cite{sed_overview,sedl_crnn,joint_sedl} and speaker diarization~\cite{diarization_overview,av_trans,beamformit}.
Currently, these tasks are primarily approached via deep-learning classifiers operating on spectrograms, though some methods do leverage localizing information based on the assumption that different audio sources are going to be in different locations~\cite{sedl_crnn,joint_sedl,av_trans,beamformit}.

Acoustic Source Localization~\cite{ssl_review,locata} is itself also an entire research topic with a rich literature, with the goal being to identify the location of a point source from which audio has been collected, or potentially just the direction relative to a reference point.
This task is generally performed using multiple microphones, with microphone arrays primarily leveraging time difference of arrival between the heavily-correlated array channels, and distributed microphones typically leveraging the varying signal amplitudes~\cite{ssl_review}.

Some information about source location is nevertheless captured by a single microphone, given sufficient supplementary information~\cite{panel_doa,mono_dist_est}.
For example, microphone directionality can be used to identify how on-axis an acoustic source is, as long as the expected loudness of the source is known.
In this work, we aim to use reverberation for the extra information necessary to at least discriminate between source locations.
To develop the intuition behind our approach, one can imagine being in a heavily-reverberant cathedral and being able to differentiate between someone nearby talking quietly and someone faraway talking loudly, despite the perceived loudness being the same, due to the latter sounding more reverberant.

Our approach is statistical in nature and leverages the Weighted Prediction Error~(WPE)~\cite{wpe1,wpe2} method of dereverberation.
This choice is due to WPE's strong record of performance and robustness to different conditions, being the standard method of practical deverberation for over a decade~\cite{reverb_challenge,best_chime5,best_chime6,best_chime8,ntt_chime8}, even through the development of neural methods~\cite{dnn_enhancement_overview,neural_wpe,dnn_cirm_derev,deep_target_cancellation}.
This is likely in part due to the stability of the late-tail component of reverberation that WPE explicitly attempts to cancel out.
We hope that this stability also lends to being a robust source of information necessary for single-channel localization.

The approach consists of estimating a likelihood ratio of whether two sets of WPE filters were extracted from sources at the same or different locations.
Unfortunately, as WPE filters are computed on a per-frequency basis in spectral domain, the approach relies on both sources containing energy in a broad range of shared frequencies.
As a result, the approach is most amenable to speech signals.

To the best of our knowledge, our task formulation is unique.
However, speaker diarization is a well-established task with strong baselines suitable for benchmarking our method.
In testing our method on a variety of synthetic and real conditions, we find our method reasonably competitive with a system using deep-learning-based speaker identification embeddings.
And, as location and vocal quality are unrelated, the approaches have uncorrelated performance, showing potential for fusion for stronger overall performance.

\subsection{Related Work}
\label{ssec:related_work}

There has been some prior work in the domain of multi-source WPE~\cite{joint_denoise_derev_sep,ive_bss_derev,iss_derev_sep}, based on the fact that different sources will have different filters, particularly for source separation.
The presence of reverberation has also been used in this way to improve source separation using non-WPE methods~\cite{cluster_bss_derev,joint_bss_drev_nmf,bss_derev_ica,conv_pred_derev_sep,ctf_sep_derev}.
Finally, Khokhlov et al.~\cite{rir_detection} have demonstrated diarization capabilities using a deep-learning impulse response classifier, but do not test their diarization method on real conversations with moving talkers, a use case that has been shown to stress deep-learning reverberation models~\cite{dnn_enhancement_overview,reverb_challenge}.

\section{Background Theory}
\label{sec:background_theory}
\subsection{Reverberation}
\label{ssec:reverberation}

For a non-moving source, the observed signal~${ x \in \mathbb{R}^T }$ can be modeled as the convolution of the source audio~${ s \in \mathbb{R}^T }$ with the room impulse response~${ h \in \mathbb{R}^L }$, which models the acoustic propagation effects from a source to a microphone in a given room:
\begin{equation}
x(t) = h(t) * s(t) = \sum_{\tau=0}^{L-1} h(\tau)s(t-\tau)
\mathrm{.}
\end{equation}
Subject to some properties of the short-time Fourier transform~(STFT), this effect can be approximated as a ``complex'' convolution operation in the STFT domain along the time axis~\cite{assse}:
\begin{equation}
X(n, f) = \sum_{k=0}^{R-1} H^*(k, f)S(n-k, f) \label{eq:stft_conv}
\mathrm{,}
\end{equation}
where~${ X \in \mathbb{C}^{N \times F} }$ is the STFT of~$ x $, ${ S \in \mathbb{C}^{N \times F} }$ is the STFT of~$s$, ${ H \in \mathbb{C}^{R \times F} }$ is \textit{not} the STFT of $h$---but rather is an approximation of the operation in the STFT domain that is equivalent to the effect of $h$ in the time domain---and $^*$ is the complex conjugate operation.

Additionally, the impulse response can be roughly decomposed into three temporal segments described characteristically~\cite{speech_dereverberation}:
\begin{equation}
h(t) = h^\text{DP}(t) + h^\text{ER}(t) + h^\text{LT}(t).
\end{equation}
The first component, $h^\text{DP}(t)$, is the direct path from the source to the microphone, representing just attenuation and delay due to propagation.
The second component, $h^\text{ER}(t)$, is the early reflections, i.e. a composition of relatively coherent copies of the source audio that have reflected only a few times off the room surfaces, and accordingly depends very heavily on the source, microphone, and room geometry and can vary significantly in time.
The final component, $h^\text{LT}(t)$, the late tail, is the later part of the reverberation, where there have been many reflections and many interferences, and the observations are more like random noise with a spectral shape matching the source audio and a temporal envelope defined by the room geometry---but with minimal dependence on the specific source or microphone location.
This breakdown can also be applied to the STFT analogue $H$ as well.

\subsection{Monaural Weighted Prediction Error}
\label{ssec:mono_wpe}

The WPE~\cite{wpe1,wpe2} method of dereverberation aims to cancel out the late tail component of the reverberation, by assuming that it is a linear function of the earlier observations.
This is motivated both by the relative stationarity of the late tail as well as by trying to minimize potential degradation of the direct path.

Nakatani et al.~\cite{wpe1} reformulate \cref{eq:stft_conv} in auto-regressive form:
\begin{equation}
X(n, f) = E(n, f) + \sum_{k=D}^{D+K-1}G^*(k, f)X(n-k, f) \label{core_decomposition}
\mathrm{,}
\end{equation}
where~${ E \in \mathbb{C}^{D \times F} }$ captures the direct path and early reflections, ${ G \in \mathbb{C}^{K \times F} }$ are the prediction filter coefficients capturing the reverberant tail past the initial $D$ frames.
The goal of WPE is to produce an estimate $\hat{E}$, i.e. $X$ with the late tail components removed, by estimating $G$.
This estimation is ultimately performed via maximum likelihood estimation and an iterative procedure under the assumption that $E$ follows a zero-mean complex Gaussian distribution.

\section{Proposed Method}
\label{sec:proposed_method}
The proposed method is based on discriminating sources in different locations by the similarities and differences in their impulse responses, as manifested in $H_1$ and $H_2$, as captured by the WPE filters.

As the direct path impulse response $h^\text{DP}(t)$ is just delay and attenuation, if the difference in delay between two sources lies within one STFT window, the relationship can be described:
\begin{equation}
H_1^\text{DP}(f) = \alpha e^{jf\theta} H_2^\text{DP}(f)
\mathrm{,}
\end{equation}
where $\alpha$ reflects the difference in magnitude resulting from the difference in propagation distance (as well as the directionality of the source and microphone, assumed negligibly frequency-dependent), and $f\theta$ is the linear phase structure resulting from the delay that reflects the relative difference in arrival time.

The relation between both pairs of early reflections and late tail are more complicated, but they can all be approximated as being drawn from complex Gaussian distributions:
\begin{align}
H_i^\text{ER}(n, f) & \sim \mathcal{N}_\mathbb{C}(0, \eta_i(n, f)) \mathrm{,}\\
H_i^\text{LT}(n, f) & \sim \mathcal{N}_\mathbb{C}(0, \gamma(n, f))
\mathrm{,}
\end{align}
where each source $i$ has its own variance $\eta_i(n, f)$, but the variance $\gamma(n, f)$ is shared among all sources in the room.

If we assume $H_i^\text{LT}(f)$ is a linear function of $H_i^\text{DP}(f)$, as modeled by WPE, then on average the WPE filters will differ in magnitude by a factor of $\alpha$.
Unfortunately the same is not necessarily true of the delay phase factor $e^{j f\theta}$.
But, if both $H_i(f)$ represent the same source, the phase difference should be zero.
And, as the prior for phase difference should be uniform, if the estimation error is uniform, assuming linear phase will not bias the distribution.

\subsection{Log Likelihood Ratio}
\label{ssec:log_likelihood_ratio}

Our method is to, given two WPE matrices $\hat{G}_1$ and $\hat{G}_2$, compute a log likelihood ratio of if the WPE matrices were extracted from the same source versus different sources.
This is done by estimating the difference in magnitude $\hat{\alpha}$ and difference in delay $\hat{d}$, and then computing the log likelihood ratio for each of these parameters independently based on distributions estimated from some training data.

To achieve order invariance and a prior on similar distances being more likely, we assume $\log{(\hat{\alpha}_\text{diff})}$ follows a normal distribution (and for convenience assume the same for $\hat{\alpha}_\text{same}$):
\begin{equation}
\log(\hat{\alpha}_\text{same}) \sim \mathcal{N}(0, \sigma^2_\text{same}),\  \log(\hat{\alpha}_\text{diff}) \sim \mathcal{N}(0, \sigma^2_\text{diff})
\mathrm{.}
\end{equation}
For the delay estimates, with different sources there is no reason to believe any particular delay difference is likely, so we assume a uniform distribution within the STFT window.
With same-source, we assume minor movements result in a bell curve, but are constrained to finite support, and accordingly choose the von Mises distribution:
\begin{equation}
\hat{d}_\text{same} \sim \mathcal{M}(0, \kappa_\text{same}),\  \hat{d}_\text{diff} \sim \mathcal{U}(0, F)
\mathrm{.}
\end{equation}
Thus, the parameters to estimate in training are $\sigma^2_\text{same}$, $\sigma^2_\text{diff}$, and $\kappa_\text{same}$.

If we assume independence of the magnitude and delay estimates and well-calibrated likelihood functions, the overall log likelihood ratio $\mathcal{L}(\hat{G}_1, \hat{G}_2)$ is merely the sum of the magnitude and delay log likelihood ratios $\mathcal{L}^\text{mag}(\hat{\alpha})$ and $\mathcal{L}^\text{delay}(\hat{d})$.
However, these are not safe assumptions, and so we additionally use Linear Discriminant Analysis~(LDA), estimated using development data, to fuse $\mathcal{L}^\text{mag}(\hat{\alpha})$ and $\mathcal{L}^\text{delay}(\hat{d})$ into a single, zero-centered score.

\subsection{Magnitude and Delay Difference Estimation}
\label{ssec:mag_and_delay_diff_estimation}

One important aspect of this problem is that WPE filters estimated in frequencies without energy from a source will be a large source of error.
And, when comparing two sets of filters, the estimates will only be reliable in bands where both sources had energy.
We thus use a frequency-wise joint-energy weighting factor in our estimates:
\begin{gather}
\epsilon_f = \sqrt{\frac{\overline{|X_1|^2}(f)}{\sum_{f \in F}\overline{|X_1|^2}(f)} \frac{\overline{|X_2|^2}(f)}{\sum_{f \in F}\overline{|X_2|^2}(f)}} \mathrm{,} \label{eq:weights}\\
\text{where } \overline{|X|^2}(f) \coloneq \frac{1}{N} \sum_{n \in N} X(n, f) X^*(n, f)
\mathrm{.}
\end{gather}

For estimating $\hat{\alpha}$: the WPE filters should differ by the factor $\alpha$, and a weighted average of the ratio of the filters suffices:
\begin{equation}
\hat{\alpha} = \frac{1}{KF}\sum_{k\in K,\, f \in F}\epsilon_f\frac{|G_1(k,f)|}{|G_2(k,f)|} \label{eq:ratio}
\mathrm{.}
\end{equation}

Estimating delay is more difficult, as something analogous to eq.~\ref{eq:ratio} only provides frequency-wise phase observations $\omega(k, f) = \angle G_1(k,f)/G_2(k, f)$.
But, a maximum likelihood estimate $\hat{d}$ can be done assuming linear phase structure, with uniform prior on the delay $D$ and von Mises-distributed errors:
\begin{gather}\hat{d} = \argmax_d p(D = d\ |\ \boldsymbol\Omega = \boldsymbol\omega),\\
\Omega[f, k] = (2\pi D\frac{f}{F} \text{ mod } 2\pi) + M \mathrm{,}\\
D \sim \mathcal{U}\{0, F-1\},\  M \sim \mathcal{M}(0, \epsilon_f\kappa)
\mathrm{.} 
\end{gather}
Using Bayes' theorem and assuming uncorrelated errors, i.e. independent $p(\Omega[f, k]\ = \omega[f,k] \ |\ D = d)$, results in:
\begin{gather}
\hat{d} = \argmax_d \prod_{k \in K, f \in F} p(\Omega[f, k] = \omega[f, k]\ |\ D=d) \mathrm{,} \\
p(\Omega[f, k] = \omega[f, k]\ |\ D=d ) = \frac{e^{\epsilon_f\kappa \cos(\omega[f, k] - (2\pi d\frac{f}{F}\text{ mod }2\pi))}}{2\pi I_0(\epsilon_f\kappa)}
\mathrm{.}
\end{gather}
After reducing terms, and remembering that $\omega = \angle G_1/G_2 \equiv \angle G_1 G_2^*$, which can be written with in terms of the cross-correlation $\mathbf{G} \coloneq G_1 G_2^*$, the final estimate of $\hat{d}$ is:
\begin{equation}
\hat{d} = \argmax_{d} \sum_{k \in K} \frac{1}{F} \sum_{f \in F} \frac{\epsilon_f}{|\mathbf{G}[f, k]|}\mathbf{G}[f, k]e^{j 2\pi d \frac{f}{F}}
\mathrm{,}
\end{equation}
i.e. essentially the argmax of the average inverse STFT of the phase differences.
This result is noteworthy as it strongly resembles the widely-used Generalized Cross-Correlation~(GCC)~\cite{gcc} method of time difference of arrival~(TDOA) estimation in beamforming.

\section{Experimental Configuration}
\label{sec:experimental_configuration}
Although our method makes no speech-specific assumptions, we choose to evaluate our method on a speaker diarization task.
This is because the proposed approach requires wideband signals to ensure enough of the WPE filters are usable across both audio sources.
However, a downside is that a localization-based diarization approach only works if the talkers do not move around too much, which is often not a safe assumption.

\subsection{Diarization Pipeline}
\label{ssec:diarization_pipeline}

The baseline and proposed system are both implemented in a clustering-based diarization framework.
First, the speech regions are broken down into short segments via a sliding window operation, and a pairwise similarity score is computed between all segments in the recording.
In the baseline system, xvector~\cite{xvectors} speaker identity embeddings are extracted and scored using PLDA~\cite{plda}, while in the proposed system the same/different log likelihood ratio with LDA fusion is computed for each pair.
Agglomerative hierarchical clustering is then used to assign the sliding-window segments to clusters corresponding to speaker identity.
In our experiments, we cluster to the ground truth number of speakers as well as explore score-based thresholding tuned on development data.

The xvector system is a publicly-available ReNet-101~\cite{resnet} system that was trained along with the PLDA on VoxCeleb1~\cite{voxceleb1}, VoxCeleb2~\cite{voxceleb2}, and CN-Celeb~\cite{cnceleb}.
The xvectors were extracted on a window of length \SI{1.44}{s} with a shift of \SI{0.24}{s}.

We estimated WPE filters on STFT windows of length 256 with shift of 128 and the typical WPE settings of length-10 filters using a delay of 3 with 3 iterations.
The length and shift of the sliding window for WPE comparisons were varied experimentally.

For evaluation, we computed the standard Diarization Error Rate~(DER) metric.
No collar was used, since we used ground truth segmentation, and overlaps were not excluded from scoring, despite both systems being limited to exactly one output speaker.

\subsection{Datasets}
\label{ssec:datasets}

The datasets we used are a dataset called Linear WHAMR! (derived from the speech separation dataset WHAMR!~\cite{whamr}), LibriCSS~\cite{libricss}, and the AMI Meeting Corpus~\cite{ami}.
All use a \SI{16}{\kilo\hertz} sample rate.

Linear WHAMR! represents the easiest condition, with two non-overlapping sources that have been artificially reverberated using synthetic impulse responses from the same room, generated using the image method~\cite{rir}.
It was created by taking the full ground truth reverberated sources from the mixtures of WHAMR!, stripping the digital silence, and concatenating them in time.

LibriCSS is a more realistic scenario, featuring 8-speaker playback speech in a conference room.
So, the reverberation is real, but the sources are non-moving with consistent impulse responses.

Finally, AMI is the most realistic scenario, with real recordings of four people in a conference room, from which we used the single distant microphone condition.
However, the recordings are roughly 30 minutes in length, and people move around, which could be catastrophic for our method's diarization performance.
As a result, we experimentally explore splitting the recording into chunks of a shorter length, within which talkers are less likely to have moved significantly.
These are evaluated independently, rather than including any sort of stitching procedure to diarize the entire recording.

\section{Results and Discussion}
\label{sec:results_and_discussion}
\begin{table}[t]
\caption{Full breakdown of performance, DER (\%). The best-performing WPE-Loc. and xvector systems are in boldface type.}
\centering
\fontsize{8}{9}\selectfont
\sisetup{
    reset-text-series = false,
    text-series-to-math = true,
    mode=text,
    tight-spacing=true,
    round-mode=places,
    round-precision=1,
    table-format=2.1,
    table-number-alignment=center}
\begingroup
\addtolength{\tabcolsep}{-0.1em}
\begin{tabular}{ccSSSS}
\toprule

& & {\rotatebox[origin=l]{90}{\textbf{L.WHAMR!}}} & {\rotatebox[origin=l]{90}{\textbf{LibriCSS}}} & {\rotatebox[origin=l]{90}{\textbf{AMI}}} & {\rotatebox[origin=l]{90}{\textbf{AMI (\SI[round-mode=none]{30}{s})}}} \vspace{-0.5em} \\ \midrule
\multirow{5.2}{*}{\rotatebox[origin=c]{90}{\begin{tabular}[c]{@{}c@{}}ground truth\\ $n_\text{spk}$ clust.\end{tabular}}} & WPE-Loc. mag. only & 26.71 & 71.00 & 60.67 & 38.29 \\
& WPE-Loc. delay only & \bfseries 4.95 & 31.20 & 60.60 & 34.53 \\
& WPE-Loc. w/o LDA & 7.48 & 25.90 & 60.64 & 33.55 \\
& WPE-Loc.+LDA & 7.78 & 24.82 & 60.57 & 33.44 \\ \cmidrule(lr){2-6}
& xvec.+PLDA & \bfseries 3.60 & 19.36 & 33.15 & 23.55 \\ \midrule
\multirow{3.2}{*}{\rotatebox[origin=c]{90}{\begin{tabular}[c]{@{}c@{}}thresh.\\ clust.\end{tabular}}} & WPE-Loc.+LDA thresh. 0.0 & 14.78 & \bfseries 20.62 & 54.41 & \bfseries 30.31 \\
& WPE-Loc.+LDA dev-tuned & 12.05 & 20.87 & \bfseries 53.89 & 30.33 \\\cmidrule(lr){2-6}
& xvec.+PLDA dev-tuned & 5.17 & \bfseries 15.23 & \bfseries 26.54 & \bfseries 21.32 \\ \midrule
\multicolumn{2}{c}{random using ground truth $n_{spk}$} & 50.89 & 88.74 & 74.10 & 60.31 \\

\bottomrule
\end{tabular}
\endgroup
\vspace{-1.5em}
\label{tab:full_results}
\end{table}

Our core experimental results are presented in Table~\ref{tab:full_results}.
We see that the statistical-based WPE-Loc. method is relatively competitive with the deep-learning xvector method, performing well above random.

For the fully-synethetic Linear WHAMR! condition, performance is strong---only a couple percentage points behind xvectors---but breaks the overall trends in a few ways: ground truth speaker number-based clustering outperforms threshold-based in both methods, delay-only estimation outperforms the combination with magnitude, and threshold tuning shows notable gains.
All of these trends can likely be explained to some degree by the fact that the ``perfect'' synthetic impulse responses led to near-perfect phase estimates and very large $\hat{\kappa}_\text{same}$ values that lead to high-variance likelihood ratios.
The delay estimates are too good for the magnitude estimates to improve them, and LDA cannot effectively calibrate the two.

Performance on LibriCSS is very encouraging, again only lagging behind xvectors by roughly 5\% DER, despite containing 8 sources in a relatively small room with simple geometry, confirming that the approach does not only work on synthetic reverberation.

With the real speech of AMI, however, the performance is unfortunately considerably weaker, though still appreciably above random.
But, the recordings are roughly 30 minutes in length, an unreasonable amount of time to assume that people do not move.
Splitting AMI into \SI{30}{s} chunks that are processed independently closes the gap to a degree, but it nevertheless remains the worst test condition.

Two noteworthy overall trends are that the use of LDA has only a very minor positive improvement on the method, and that tuning the clustering threshold on the development set also results in minimal improvement.
Though the experiments are omitted due to space constraints, these points are actually connected---we found that the use of LDA successfully calibrates the scores such that the optimal tuning threshold is effectively 0, eliminating the need for tuning. Omitting the LDA fusion leads to threshold tuning being necessary.

\begin{table}[t]
\caption{DER (\%) performance of LDA-calibrated WPE-Loc. with a clustering threshold of 0.0 on LibriCSS.}
\centering
\fontsize{8}{9}\selectfont
\sisetup{table-format=2.1,round-mode=places,round-precision=1,table-number-alignment=center}
\begingroup
\addtolength{\tabcolsep}{-0.2em}
\begin{tabular}{cSRRRRRRRRR}
\toprule

& & \multicolumn{9}{c}{\textbf{window (s)}} \\
& & \multicolumn{1}{c}{\bfseries{1.0}} & \multicolumn{1}{c}{\bfseries{1.5}} & \multicolumn{1}{c}{\bfseries{2.0}} & \multicolumn{1}{c}{\bfseries{2.5}} & \multicolumn{1}{c}{\bfseries{3.0}} & \multicolumn{1}{c}{\bfseries{3.5}} & \multicolumn{1}{c}{\bfseries{4.0}} & \multicolumn{1}{c}{\bfseries{4.5}} & \multicolumn{1}{c}{\bfseries{5.0}} \\ \cmidrule(lr){3-11}
\multirow{5}{*}{\rotatebox[origin=c]{90}{\textbf{shift (s)}}} & \bfseries{0.5} & 55.1 & 37.7 & 29.0 & 24.2 & 21.7 & 20.8 & 20.6 & 21.6 & 22.7 \\
& \bfseries{1.0} & \multicolumn{1}{c}{} & 39.5 & 30.2 & 25.2 & 22.6 & 20.9 & 20.8 & 22.1 & 22.0 \\
& \bfseries{1.5} & \multicolumn{1}{c}{} & \multicolumn{1}{c}{} & 30.4 & 25.4 & 23.0 & 21.4 & 21.4 & 21.9 & 32.2 \\
& \bfseries{2.0} & \multicolumn{1}{c}{} & \multicolumn{1}{c}{} & \multicolumn{1}{c}{} & 26.2 & 22.9 & 21.9 & 36.3 & 35.3 & 36.5 \\
& \bfseries{2.5} & \multicolumn{1}{c}{} & \multicolumn{1}{c}{} & \multicolumn{1}{c}{} & \multicolumn{1}{c}{} & 24.0 & 34.4 & 37.4 & 37.1 & 37.2 \\

\bottomrule
\end{tabular}
\endgroup
\vspace{-1.5em}
\label{tab:slwin}
\end{table}

Table~\ref{tab:slwin} contains the results of our experiments tuning the extraction window and shift used for WPE-Loc.
Unfortunately, as expected, the proposed approach requires a much larger extraction window compared to the xvector system.
The resulting window and shift of \SI{4.0}{s} and \SI{0.5}{s} was used for all other experiments.

\begin{figure}[tbh]
\centering
\includegraphics[width=\columnwidth]{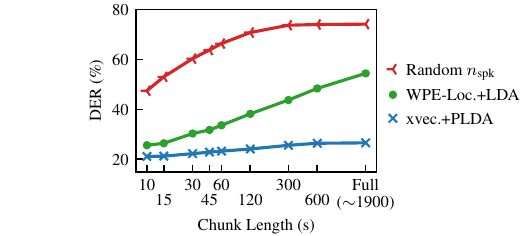}
\vspace{-16pt}
\caption{Performance on AMI when the recordings are segmented into equal-length chunks that are processed and scored independently, to mitigate moving-source errors. The WPE-Loc.+LDA system uses a clustering threshold of 0.0; xvec.+PLDA uses a threshold of -0.15.}
\vspace{-4pt}
\label{fig:chunk_plot}
\end{figure}

We also further investigated the division of AMI into chunks to minimize moving-source effects, the results of which are presented in Figure~\ref{fig:chunk_plot}.
As expected, the xvector system is fairly invariant to this, but with WPE-Loc. the error rate drops with chunk length---getting within 5\% of xvectors---with a log-linear relationship over the full span from \SI{10}{s} to \SI{30}{m}, as should be expected if movements occur randomly and are proportional in count to the chunk length.
The fact that WPE-Loc. performance immediately begins improving as chunk size is decreased---well before the random baseline starts improving due to fewer sources per chunk---indicates that this performance improvement is likely due to the issue of sources moving rather than due to shorter chunks resulting in an easier task.

\begin{figure}[tbh]
\centering
\includegraphics[width=\columnwidth]{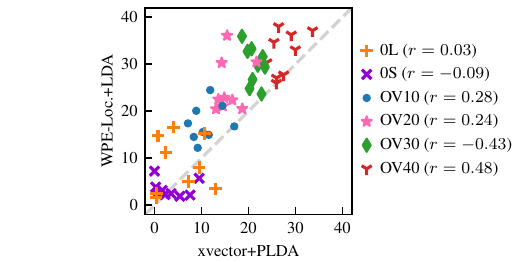}
\vspace{-16pt}
\caption{Recording-wise DER (\%) comparison of the top-performing WPE-Loc. and xvector systems on LibriCSS, categorized by overlap condition (0L and 0S contain no overlap, OV$n$ contains $n$\% overlap), along with the correlation coefficient between the methods.}
\label{fig:correlation}
\end{figure}

Finally, we wanted to confirm that the two methods are picking up on different cues by comparing per-recording performance on LibriCSS, as shown in Figure~\ref{fig:correlation}.
Unsurprisingly, the amount of overlap is the strongest predictor of performance, but when normalizing for the amount of overlap in the recording, there is almost no correlation between the two methods, suggesting high potential for fusion.
In the no-overlap condition, we even see that almost half the time, the WPE-Loc. method actually outperforms the xvector system.
Performance is still close with overlap, however, and the relative degradation of WPE-Loc. might be due to the larger extraction window leading to more representations corrupted by overlap.

\section{Conclusion}
\label{sec:conclusion}
We have developed a statistical framework for discriminating acoustic sources in different locations based on WPE dereverberation filters.
Experimental results have shown that the proposed method can achieve performance close to a deep learning speaker-ID system on speaker diarization, while cueing on different information, showing promise for improving state-of-the-art speech temporal segmentation tasks.

\vfill\pagebreak

\bibliographystyle{IEEEtran}
\bibliography{strings}

\end{document}